# Impact of initial outbreak locations on transmission risk of infectious diseases in an intra-urban area

**Kang Liu**
Kang.liu@siat.ac.cn
Shenzhen Institute of Advanced Technology, Chinese Academy of Sciences
**Ling Yin**
yinling@siat.ac.cn
Shenzhen Institute of Advanced Technology, Chinese Academy of Sciences
**Jianzhang Xue**
psionic@mail.ustc.edu.cn
University of Science and Technology of China

## Abstract

Infectious diseases usually originate from a specific location within a city. Due to the heterogenous distribution of population and public facilities, and the structural heterogeneity of human mobility network embedded in space, infectious diseases break out at different locations would cause different transmission risk and control difficulty. This study aims to investigate the impact of initial outbreak locations on the risk of spatiotemporal transmission and reveal the driving force behind high-risk outbreak locations. First, integrating mobile phone location data, we built a SLIR (susceptible-latent-infectious-removed)-based meta-population model to simulate the spreading process of an infectious disease (i.e., COVID-19) across fine-grained intra-urban regions (i.e., 649 communities of Shenzhen City, China). Based on the simulation model, we evaluated the transmission risk caused by different initial outbreak locations by proposing three indexes including the number of infected cases (CaseNum), the number of affected regions (RegionNum), and the spatial diffusion range (SpatialRange). Finally, we investigated the contribution of different influential factors to the transmission risk via machine learning models. Results indicates that different initial outbreak locations would cause similar CaseNum but different RegionNum and SpatialRange. To avoid the epidemic spread quickly to more regions, it is necessary to prevent epidemic breaking out in locations with high population-mobility flow density. While to avoid epidemic spread to larger spatial range, remote regions with long daily trip distance of residents need attention. Those findings can help understand the transmission risk and driving force of initial outbreak locations within cities and make precise prevention and control strategies in advance.

# 1 Introduction

Over the past decades, the outbreaks of chronic or emerging infectious diseases have posed a growing threat on public health, global economy, and human society (Bloom and Cadarette, 2019). Many infectious diseases like COVID-19 are transmitted from individual to individual through physical contacts. Such transmission is often fostered and accelerated in cities with dense populations, well-developed transportation, and high-dynamic human mobility (Mao and Bian, 2010). Reviewing the historical and current epidemics, it can be found that infectious diseases breaking out in cities usually originated from a specific location, and then spread to a larger spatial range. For example, the COVID-19 epidemic in Wuhan was initially concentrated in Huanan Seafood Wholesale Market at the end of 2019 (Li et al., 2020a); the second wave COVID-19 epidemic in Beijing starting from June 2020 was initially concentrated in the Xinfadi Agricultural Products Wholesale Market, and then spread to broader space of the city (Wei et al., 2020).

Due to the differentiation in functional planning and self-development, regions within a city usually have significant differences in population and public facilities. Such spatial heterogeneity further results in heterogeneity of the structure of human mobility network embedded in space (Liu et al., 2020; Hou et al., 2021). In such situation, infectious diseases break out at different locations would generate disparate spatiotemporal transmission risk and cause different degree of difficulty in prevention and control. Therefore, it is of great importance to explore the influence of initial outbreak locations on transmission risk and investigate the driving factors behind high-risk outbreak locations.

To quantificationally and systematically study the problem of initial outbreak locations, it is essential to depict the spread of infectious diseases in time and space by mathematics and computer modeling. Researchers have proposed a large number of infectious-disease transmission models, at both the group level (Chen et al., 2020; Gatto et al., 2020; Lai et al., 2020; Li et al., 2020b; Hou et al., 2021; Huang et al., 2021a; Yang et al., 2020; Zhou et al., 2020) and the individual level (Aleta et al., 2020; Hellewell et al., 2020; Koo et al., 2020; Yin et al., 2021).

The group-level models are mainly implemented by differential equations, which can be built for the whole study area, or for each of the sub-regions, respectively. For the latter type, the models of sub-regions would be coupled together based on human mobility between them. Such coupled models are called meta-population models, which have become one of the most popular model types at present. For instance, by integrating human mobility data collected by the Tencent Location-based Service, Li et al. (2020b) developed an SEIR (susceptible-exposed-infectious-removed)-based meta-population model to simulate the spatiotemporal dynamics of the infections in 375 Chinese cities. This study found that substantial mildly and asymptomatically undocumented infections facilitate the rapid dissemination in the early stage of COVID-19 epidemic in China. Lai et al. (2020) formed a similar meta-population model for 340 cities in mainland China and evaluated the effects of three types of NPI (non-pharmaceutical intervention) executed in China after the COVID-19 outbreak. Zhou et al. (2020) built a SEIR-based meta-population model for 10 administrative divisions of Shenzhen City, China based on mobile phone location data to quantitatively assess the effectiveness of four types of travel restrictions.

The individual-level models are usually implemented by agent-based models, in which individuals are designed as agents with demographic attributes and infection status. The interactions between agents determine the dynamic diffusion process of infectious diseases (Hunter et al., 2017). For instance, Koo et al. (2020) constructed an agent-based epidemic simulation model for a synthetic population of Singapore derived from traffic data and census data, and then evaluated the effectiveness of four intervention scenarios for COVID-19. Combining mobile phone location data and census data, Albert et al. (2020) created an agent-based model in the Boston metropolitan area for COVID-19. Based on the simulation results of the model, they recommended an acceptable trade-off between the reopening of economic activities and the capacity of the healthcare system. Yin et al. (2021) integrated massive mobile phone location data, census data and building characteristics into a spatially explicit agent-based model to simulate COVID-19 spread among 11.2 million individuals living in Shenzhen City. Based on the model, they assessed the probability of COVID-19 resurgence if sporadic cases occurred in a fully reopened city.

Even though a variety of studies have been made based on those epidemiological models, most studies have focused on the prediction of spatiotemporal transmission trends (Yang et al., 2020) or the simulation and assessment of various NPIs (Aleta et al., 2020; Chinazzi et al., 2020; Koo et al., 2020; Kraemer et al., 2020) at spatial scales of countries, provinces/states, and cities (Bajardi et al., 2011; Li et al., 2020b; Lai et al., 2020; Yang et al., 2020). Few studies have delved into a fine-grained intra-urban scale and investigated the impact of different initial outbreak locations on spatiotemporal transmission risk of infectious diseases. To fill the research gap, this study aims to systematically quantify, compare, and analyze the transmission risk caused by different initial outbreak locations and reveal the driving force behind the high-risk initial outbreak locations. To achieve this goal, we firstly built a SLIR-style meta-population model to simulate the transmission process of COVID-19 across 649 communities in Shenzhen City based on mobile phone location data and point-of-interest (POI) data. Then, we proposed three risk evaluation indexes (i.e., the number of infected cases, the number of affected regions, and the spatial diffusion range) to assess the transmission risk caused by different initial outbreak sites. Finally, taking the risk evaluation indexes as dependent variables, we applied random forest models to compare the contribution of potential influential factors (e.g., population and human mobility flow) to the transmission risk. Our study can not only help understand the transmission mechanism of infectious diseases, but also assist governments to develop forward-looking and precise prevention and control strategies.

## 2 Study area and data

### 2.1 Study area

Shenzhen is one of the most developed metropolitan cities in China, which encloses a permanent population of 17.56 million[1] in an area of 1997.47 km$^2$. Considering the high population density and human mobility dynamic, we selected Shenzhen as a typical city for the case study. Figure 1 shows the 649 communities of Shenzhen City. The spatiotemporal transmission process was simulated at such fine spatial scale, and each of the communities was served as an initial outbreak location in the case study.

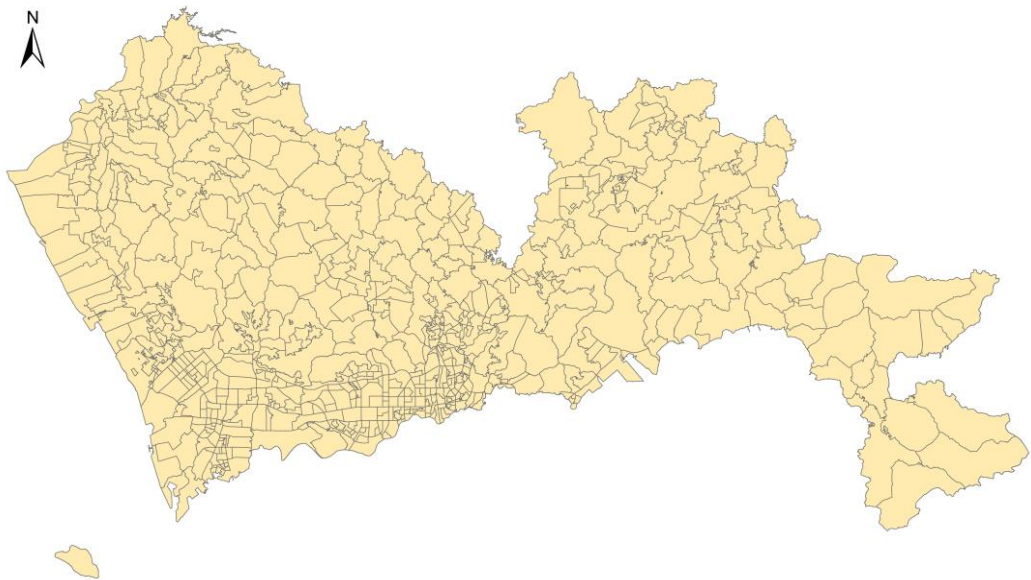

**Figure 1. The 649 communities of Shenzhen City, China.**

### 2.2 Mobile phone location data

The mobile phone location data used in the study were provided by one of the three major telecommunication operators in China, China Unicom, which holds the location records of more than 400 million subscribers in China. An original location record includes the user's anonymous ID, and the timestamp and coordinate of the cell tower she/he was connected. Furthermore, we entrusted SmartSteps[2], a company hold by China Unicom, to process the original mobile phone location data to several types of data, including residential population, working population, daily population flow, and stay-all-day population. Since China Unicom only covers part of the city population, SmartSteps made sample expansions according to their market shares in Shenzhen. The total population of the city after sample expansion is 16.61 million, which is close to the census data of 17.56 million.

---

[1] http://tjj.sz.gov.cn/ztzl/zt/szsdqcqgrkpc/szrp/content/post_8772112.html

The definitions and processing methods of residential population, working population, daily population flow, and stay-all-day population are described as follows.

1) Residential population

    The residential community of an individual is defined as the community where the individual spent the longest time during the nighttime (21:00 to 07:00) over a month (i.e., October 2019 in our study case). Based on the definition, the residential population of each community can be acquired. To ensure that the individual is a long-stay resident instead of a short-stay visitor, we only counted individuals who had appeared in the residential community for more than two weeks (14 days) in the month.

2) Working population

    The working community of an individual is defined as the community where the individual spent the longest time during the daytime (09:00 to 17:00) over a month (i.e., October 2019 in our study case). Based on the definition, the working population of each community can be acquired.

3) Daily population flow

    The daily population flow indicates home-based daily population mobility flow between communities. Figure 2 demonstrates the OD-based trips and home-based trips of an individual during a day. An OD-based trip is defined as a trip from one location (i.e., origin) to the next location (i.e., destination), while a home-based trip is defined by setting the home location as the origin, and each of the locations that the individual has visited during the day as the destination. On this basis, daily population flow is defined as the total number of home-based trips from one community to the other during a day. We further averaged the daily population flow data of 14 days (i.e., October 14

---

[2] http://www.smartsteps.com/

to 27, 2019) for the use of the case study.

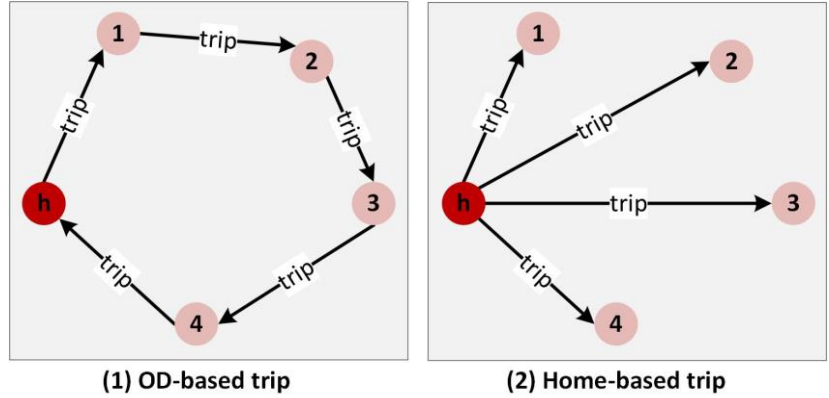

**Figure 2. OD-based trips and home-based trips of an individual during a day.**

4) stay-all-day population

The stay-all-day residential population of a community was defined as the number of residents that did not leave in their residential communities during a day. Similarly, we averaged the stay-all-day population data of 14 days (i.e., October 14 to 27, 2019) for the use of the case study.

**2.3 Point-of-interest data**

POIs depict places where people gather and conduct their daily activities in a fine spatial and taxonomic granularity. In this study, we collected about 1.7-million POIs of Shenzhen City from the Open Platform of AutoNavi[3], one of the largest web mapping, navigation, and location-based services providers in China. Each POI includes information of POI name, type, coordinate, and address.

All the POIs were spatially joined with the communities by the ArcGIS software, and then we counted the number of POIs inside each community for the use of the case study.

# 3 Modeling intra-urban spatiotemporal transmission process of COVID-19

Taking COVID-19 as the typical infectious disease, we built a SLIR-based meta-population model to simulate the spatiotemporal transmission across fine-grained intra-urban regions (i.e., the 649 communities in Shenzhen in our study case).

For community $i$, its residential population $N_i$ is divided into susceptible population $S_i$, latent population $L_i$, infectious population $I_i$ (including symptomatic infectious population $I_i^1$ and asymptomatic infectious population $I_i^0$) and removed population $R_i$ (including population with recovered, quarantined, and dead statuses), respectively.

---

[3] https://lbs.amap.com/api/webservice/guide/api/search

According to the epidemiological characteristics of COVID-19 in the early stage of its outbreak, 75% (Koo et al., 2020; Mizumoto et al., 2020; Nishiura et al., 2020) of the infected cases had obvious symptom after an incubation period of a mean of 5.2 days (Lauer et al., 2020; Li et al., 2020a). The symptomatic cases became infective 1-2 days before symptom onset (WHO and China, 2020) and were hospitalized about 6 days after symptom onset (Li et al., 2020a; Huang et al., 2021a; Huang et al., 2020). Therefore, we set the latent period and the infectious period of the symptomatic cases as 3.7 days (5.2-1.5 days) and 7.5 days (6+1.5 days), respectively. While 25% of the infected cases were asymptomatic throughout the course of the disease and remained infective for 9.5 days (Hu et al., 2020) after the latent period. The infectivity of the asymptomatic cases was set as 0.12 of the symptomatic cases according to the report of Chinese Center for Disease Control and Prevention (China CDC) (China CDC, 2020).

Due to the high-dynamic human mobility in an intra-urban area, susceptible individuals may be infected not only in their residential communities, but also in other communities. Therefore, by integrating the human mobility information between communities, the SILR-based meta-population model for community $i$ was built as follows:

$$\frac{dS_i(t)}{dt} = -\sum_j S_i(t) h_{ij} \beta_j \frac{\sum_l h_{lj} (I_l^1(t) + \varepsilon I_l^0(t))}{\sum_l h_{lj} N_l}$$

$$\frac{dL_i(t)}{dt} = \sum_j S_i(t) h_{ij} \beta_j \frac{\sum_l h_{lj} (I_l^1(t) + \varepsilon I_l^0(t))}{\sum_l h_{lj} N_l} - \delta L_i(t)$$

$$\frac{dI_i^1(t)}{dt} = \sigma \delta L_i(t) - \gamma_1 I_i^1(t)$$

$$\frac{dI_i^0(t)}{dt} = (1-\sigma) \delta L_i(t) - \gamma_0 I_i^0(t)$$

$$\frac{dR_i(t)}{dt} = \gamma^1 I_i^1(t) + \gamma^0 I_i^0(t)$$

The model is modified based on the one proposed by Wu et al. (2007). The parameters in the formula are described in Table 1. Except for $\beta$, all the parameter values were determined according to existing literature mentioned above. In addition, $h_{ij}$ indicates the daily population flow from community $i$ to $j$. If $i = j$, $h_{ij}$ denotes the stay-all-day population of community $j$. $h_{ij}$ was further normalized to ensure that $\sum_j h_{ij} = 1, \forall i$.

Table 1. Parameters in the SLIR-based meta-population model.

| | Description | Value |
|---|---|---|
| $\beta$ | Effective infection rate of symptomatic cases | ? |
| $\varepsilon$ | Ratio of infectivity of asymptomatic to symptomatic cases | 0.12 |
| $\delta$ | Transformation rate of latent period to infectious period | 1/3.7 days |
| $\sigma$ | Proportion of symptomatic cases | 75% |
| $\gamma^1$ | Removed rate of symptomatic cases | 1/7.5 days |
| $\gamma^0$ | Removed rate of asymptomatic cases | 1/9.5 days |

Considering that physical contacts and viral transmission between individuals usually occur in various activity places such as restaurants, cinemas, and markets, we assumed that communities with higher density of activity places would have larger effective infection rate. Therefore, the effective infection rate $\beta(j)$ of community $j$ varies with its inside POI density and can be written as:

$$\beta(j) = \beta \times \frac{POI\_density(i) - min(POI\_density(\forall i))}{max(POI\_density(\forall i)) - min(POI\_density(\forall i))},$$

where $\beta$ is a basic effective infection rate of the city.

We used the following steps to estimate the only unknown parameter $\beta$.

1) Given a specific value of $\beta$, we obtained the epidemic curve of daily new infections (including symptomatic and asymptomatic cases) of the city by simulations using the model.

2) R0 of the epidemic curve was then estimated using the exponential-growth (EG) method implemented by the "R0 package" of R programming language (Obadia et al., 2012).

3) Through the above two steps, we can obtain multiple pairs of $(\beta, R0)$. By fitting the relationship of $\beta$ and R0, we can obtain the corresponding $\beta$ value by setting R0 as 2.5 (WHO and China, 2020).

In our case, to generate the epidemic curves, 100 initial symptomatic cases were seeded in various communities approximately proportional to their population sizes. As shown in Figure 3, the linear relationship between $\beta$ and R0 is well fitted with the coefficient of determination $R^2$ up to 0.999. When R0=2.5, the $\beta$ value is equal to 0.405.

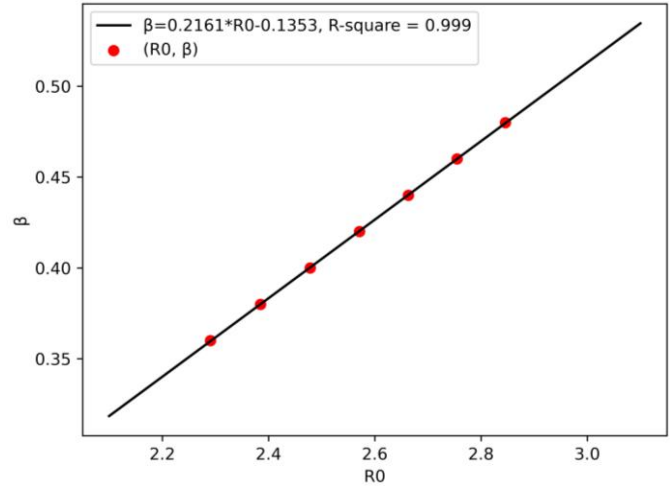

**Figure 3. Linear relationship between** $\beta$ **and R0.**

## 4 Evaluating transmission risk caused by different initial outbreak locations

To evaluate the transmission risk caused by different initial outbreak locations, we put 100 initial cases at each community, and simulated the spatiotemporal transmission process of COVID-19 using the built SLIR-based meta-population model. Similarly, 100 initial cases were seeded in various communities approximately proportional to their population sizes.

We defined three indexes to evaluate the spatiotemporal transmission risk based on the simulation results and analyzed the transmission risk caused by different initial outbreak locations.

**4.1 Three indexes for evaluating transmission risk**

We defined the following three indexes to evaluate the spatiotemporal transmission risk from different aspects.

1) **Number of infected cases (CaseNum).** Specifically, it indicates the cumulative number of infected cases of the whole city in the $i$-th day after the outbreak.
2) **Number of affected regions (RegionNum).** Specifically, it indicates the number of regions (i.e., communities in our study case) that have appeared infected cases in the $i$-th day after the outbreak.
3) **Spatial diffusion range (SpatialRange).** Specifically, it measures the geographical spatial range that has affected by infected cases in the $i$-th day after the outbreak. As shown in Figure 4, different initial outbreak locations may result in very different spatial diffusion range, even though they affected the same number of regions. To measure the spatial diffusion range, we first calculated the barycentric coordinates of the appeared infected cases across the city in the $i$-th day after the outbreak:

$$CaseCenter = \frac{\sum_1^N (CaseNum(i) \times LocCenter(i))}{\sum_1^N CaseNum(i)}.$$

Then, we calculated the spatial diffusion range:

$$SpatialRange = \frac{\sum_1^N [[dist(LocCenter(i), CaseCenter)]^2 \times CaseNum(i)]}{\sum_1^N CaseNum(i)}.$$

$N$ is the number of intra-urban regions (i.e., 649 communities); $CaseNum(i)$ is the cumulative number of infected cases in community $i$, therefore $\sum_1^N CaseNum(i)$ is the cumulative number of infected cases in the whole city; $LocCenter(i)$ is the coordinates of the centre point of community $i$; $dist(\cdot,\cdot)$ indicates the geographic distance between the given two points.

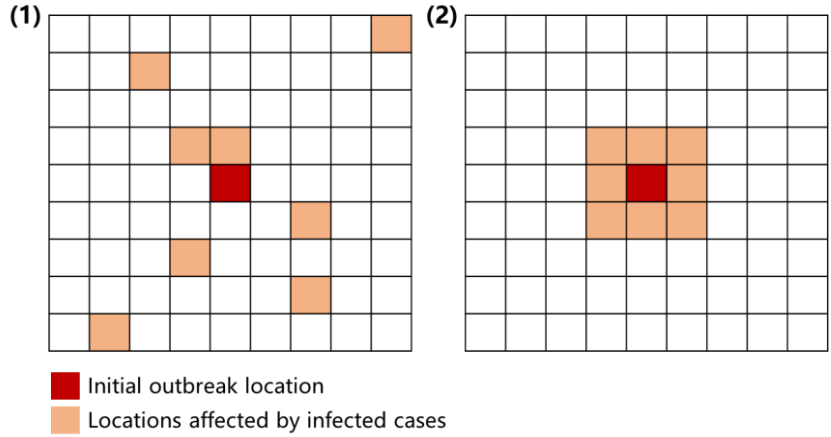

**Figure 4. Demonstration of infectious-disease transmission with the same number of infected regions (RegionNum) but different spatial diffusion range (SpatialRange).**

**4.2 Analyzing transmission risk caused by different initial outbreak locations**

By putting 100 initial infected cases (i.e., seeds) in each of the communities, respectively, we simulated the spatiotemporal transmission process of COVID-19 in Shenzhen and evaluated the transmission risk caused by different communities using the three indexes defined above.

Figure 5 shows the transmission risk evaluated by the index of "number of infected cases (CaseNum)". It indicates that infectious diseases breaking out in different locations would cause similar total number of cases in the city. This conclusion is similar to a study on influenza made by Mao and Bian (2010), where they found that in four different seeding scenarios, the epidemic curve develops similarly in the peak time, peak number of new cases, and total number of cases.

Figure 6 shows the transmission risk measured by the index of "number of infected regions (RegionNum)". It indicates that infectious diseases breaking out in different locations would affect different number of regions, especially in the early stage of the epidemic. Figure 6 (1) and (2) imply that population concentrated and developed regions would result in higher transmission risk; the more specific driving factors behind those high-risk initial outbreak locations will be investigated in section 5.

Figure 7 shows the transmission risk assessed by the index of "spatial diffusion range (SpatialRange)". It indicates that epidemic breaking out in remote locations may result in larger spatial diffusion range compared to that breaking out in geographically central locations. We suspected that this is related to the spatial heterogeneity of human travel behavior that residents living in central regions are more accessible to various places (e.g., workplaces and shopping malls), so they have shorter travel distances and less probability to bring the virus to distant places. On the contrary, residents living in remote regions usually need long-distance travel to satisfy their daily or specific needs, so they are more likely to transmit the disease to larger spatial range.

The above analyses are mainly based on the risk maps shown in Figure 4-6 as well as the authors' knowledge on Shenzhen City as residents. In section 5, we will further investigate the potential influential factors behind high-risk initial outbreak locations quantificationally.

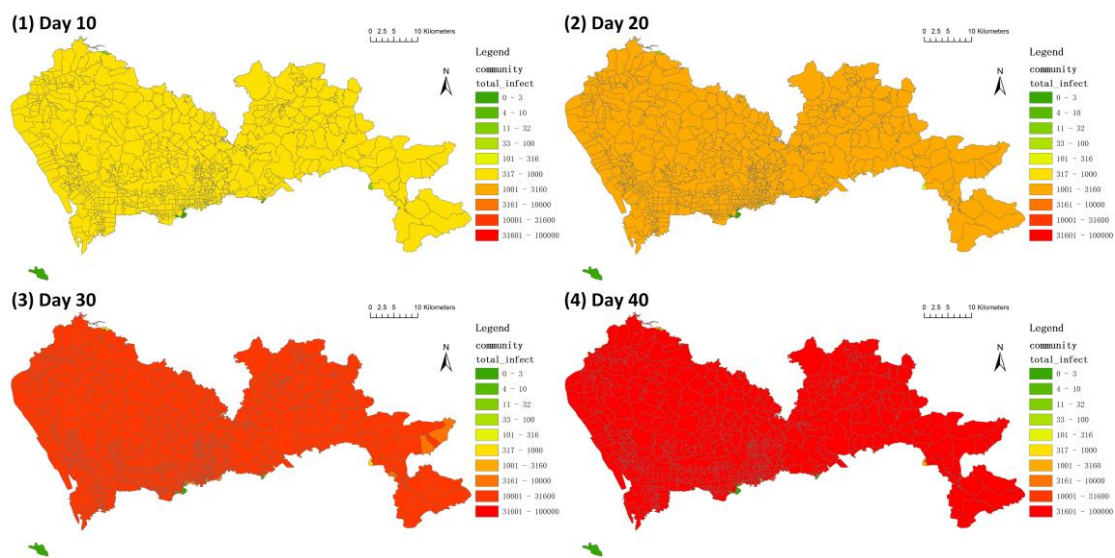

**Figure 5. Number of infected cases (CaseNum) caused by each of the 649 communities in Shenzhen as initial outbreak location.**

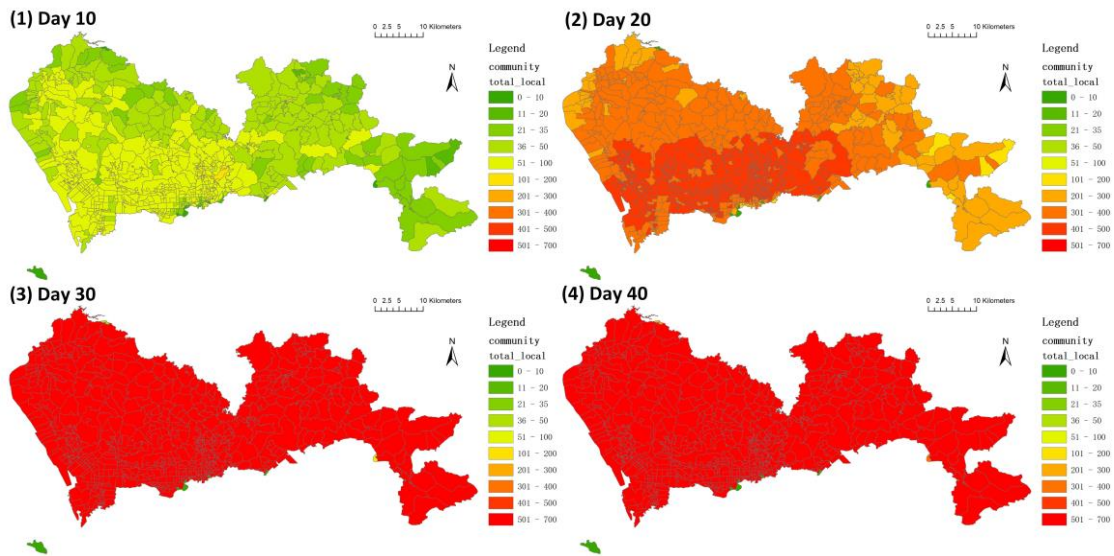

**Figure 6. Number of infected regions (RegionNum) caused by each of the 649 communities in Shenzhen as initial outbreak location.**

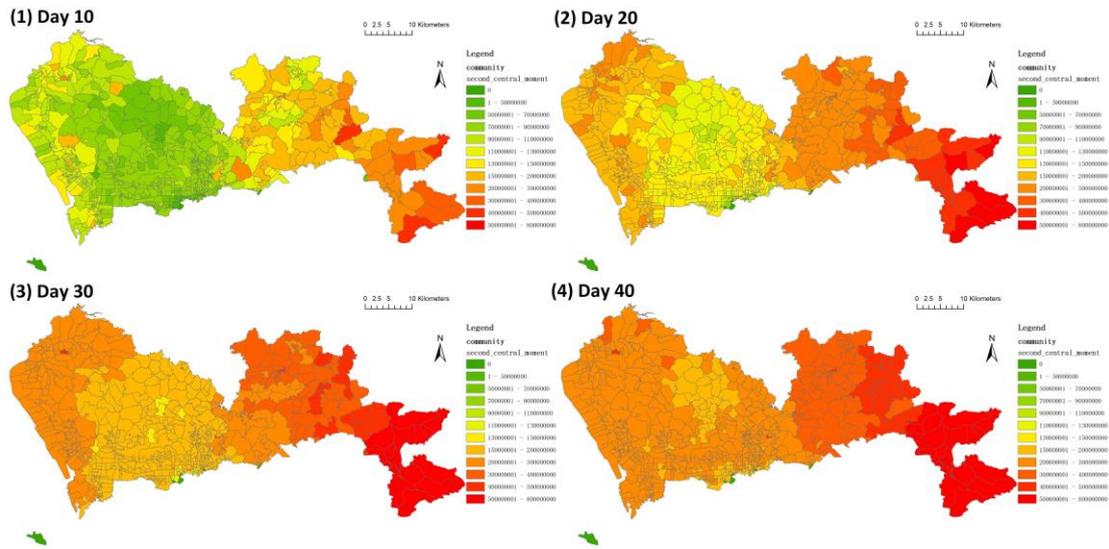

**Figure 7. Spatial diffusion range (SpatialRange) caused by each of the 649 communities in Shenzhen as initial outbreak location.**

## 5 Revealing influential factors behind high-risk initial outbreak locations

In this section, we established regression models to reveal the driving factors behind high-risk initial outbreak locations. The dependent variables of the models are the three indexes (i.e., CaseNum, RegionNum, and SpatialRange), respectively. While the independent variables include population related factors, population-flow related factors, and POI related factors. The detailed 16 independent variables are presented in Table 2.

**Table 2. Number and descriptions of the independent variables (influential factors).**

| No. | Independent variables | Descriptions |
|---|---|---|
| 1 | residential population | |
| 2 | residential population density | |
| 3 | working population | |
| 4 | working population density | |
| 5 | daily flow volume | daily inflow and outflow volume |
| 6 | daily flow density | |
| 7 | average travel distance of residents | average travel distance of all home-based daily trips of the residents |
| 8 | POI quantity | |
| 9 | POI density | |
| 10 | degree (unweighted) | Two types of population flow networks are built with each community served as a node. If there exist daily population flow between two communities, then a directed edge is built between them. In the unweighted population flow network, the weights of the edges are set as 1. While in the weighted population flow network, the weights of the edges are set as the population flow volume between the two communities. All the centrality measures were calculated using the Python package "NetworkX"[4]. |
| 11 | betweenness (unweighted) | |
| 12 | closeness (unweighted) | |
| 13 | PageRank (unweighted) | |
| 14 | betweenness (weighted) | |
| 15 | closeness (weighted) | |
| 16 | PageRank (weighted) | |

Considering that the independent variables may have multicollinearity, and the relationship between the independent variables and dependent variables might not be linear (Huang et al., 2021b), we chose random forest regressor (Liaw and Wiener, 2002), an ensemble learning technique composed of multiple decision trees, as the regression model. In addition, the random forest model can give the relative importance value of each independent variable, which is computed as the (normalized) total reduction of the model performance (i.e., criterion measured by mean squared error) brought by the variable.

As for the experiments, we divided the 649 communities of Shenzhen City into a training set (75%) and a testing set (25%) and used 100 decision trees in the random forest regressor. Pearson correlation coefficient (r) and coefficient of determination

---

[4] https://networkx.org/documentation/stable/reference/algorithms/centrality.html

($R^2$) were used to evaluate the performance of the models. Table 3 and 4 show the averaged r and $R^2$ of each model for executing 100 times. It indicates that the performance of the model is quite well.

**Table 3. Model performance evaluated by Pearson correlation coefficient (r).**

|  | **CaseNum** | **RegionNum** | **SpatialRange** |
|---|---|---|---|
| **Day 10** | 0.984 | 0.806 | 0.899 |
| **Day 20** | 0.975 | 0.911 | 0.872 |
| **Day 30** | 0.971 | 0.965 | 0.856 |
| **Day 40** | 0.963 | 0.967 | 0.844 |

**Table 4. Model performance evaluated by coefficient of determination ($R^2$).**

|  | **CaseNum** | **RegionNum** | **SpatialRange** |
|---|---|---|---|
| **Day 10** | 0.967 | 0.645 | 0.809 |
| **Day 20** | 0.951 | 0.829 | 0.760 |
| **Day 30** | 0.942 | 0.931 | 0.731 |
| **Day 40** | 0.927 | 0.936 | 0.713 |

Figure 8 shows the relative importance values of the independent variables.

As for the transmission risk evaluated by the number of infected cases (CaseNum), the relative importance of the variables is similar in different days. It indicates that factors such as residential population (No. 1), daily flow volume (No. 5), and degree (No. 10) show higher importance compared to others. However, considering that the CaseNum caused by different initial outbreak locations are similar (Figure 5), those "important factors" have limited enlightenment for prevention and control in practice.

As for the transmission risk evaluated by the number of infected regions (RegionNum), the relative importance of the variables has a change pattern with time. In the early stage (e.g., day 10 and 20), daily flow density (No. 6) plays the most important role, while in the latter stage (e.g., day 30 and 40), residential population (No. 1) and some network centrality measures of population flow network (e.g., degree and closeness) show high importance than other factors. It implies that to avoid the epidemic spreading quickly to more other regions, it is necessary to prevent the epidemic breaking out in locations of high daily flow density or make efforts to control such locations immediately after outbreak.

As for the transmission risk evaluated by the spatial diffusion range (SpatialRange), average travel distance of residents plays significant roles in different days, which implies that epidemic breaking out in locations with longer trip distance would spread to larger spatial range. Figure 7 also implies that those locations are usually suburban areas. Residents living in such regions usually need long-distance travel to satisfy their daily needs, so they are more likely to transmit the disease to larger spatial range. Therefore, to avoid the epidemic spreading to larger spatial range, the remote regions with long daily travel distance of residents also need attention.

From Figure 8, we can also find that the population density (including residential and working), POI quantity and density, and some centrality measures of mobility network (e.g., betweenness of weighted mobility network) do not show obvious influence on transmission risk compared to those mentioned above.

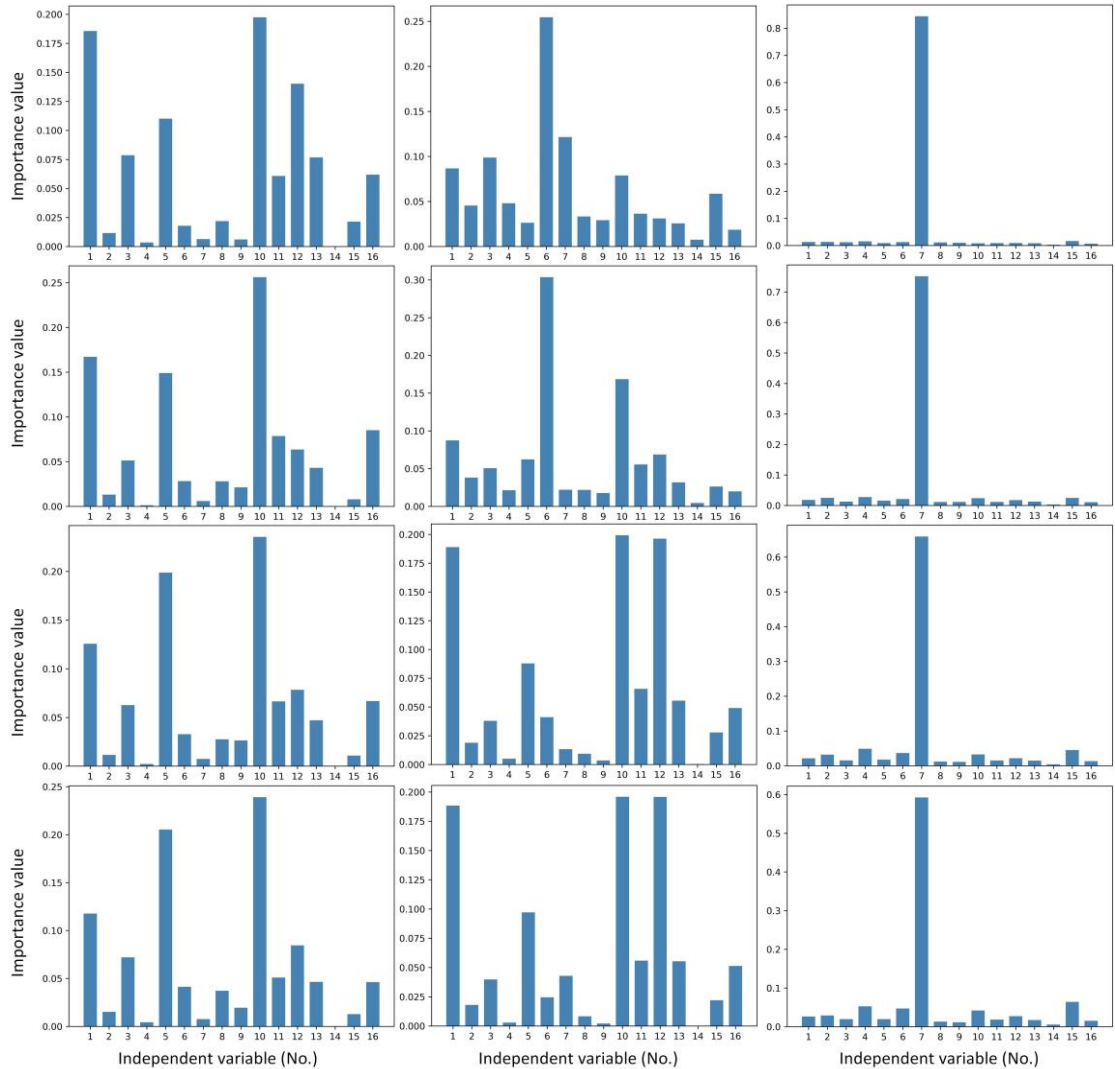

**Figure 8. Relative importance of independent variables based on the random forest regressor.**

## 6 Discussion and conclusion

Infectious diseases breaking out within a city usually originated from a specific location, and then spread to a larger spatial range. As the distributions of demographic, geographic, and socioeconomic elements have great spatial heterogeneity, infectious diseases breaking out at different locations would cause different levels of transmission risk. Therefore, it is necessary to systematically investigate the influence of initial outbreak locations on transmission risk of infectious diseases in a quantitative way. To achieve this goal, we firstly built a SLIR-based meta-population model to simulate the spatiotemporal spreading process of COVID-19, and then analyzed the transmission risk caused by each community of Shenzhen as initial outbreak location using three indexes including the number of infected cases (CaseNum), the number of affected regions (RegionNum), and the spatial diffusion range (SpatialRange). Finally, the relative impact of potential influential factors on the transmission risk was given by the random forest regressors.

We have some findings based on the experiment results.

(1) Different initial outbreak locations would cause similar CaseNum but different RegionNum and SpatialRange.

(2) Initial outbreak locations with higher daily flow density would cause larger RegionNum. To avoid the epidemic spread quickly to more regions, it is necessary to prevent the epidemic breaking out in such locations.

(3) Initial outbreak locations with longer daily trip distance of residents (those locations are usually suburbs of the city) would cause larger SpatialRange. Therefore, to avoid the epidemic spread to larger spatial range, suburban areas also require attention.

Those findings can not only help understand the mechanism of transmission risk and driving force behind high-risk initial outbreak locations, but also help make precise prevention and control strategies in advance.

The highlights or contributions of this study can be summarized as follows.

(1) On the topic of infectious-disease transmission, most studies have focused on predicting the spatiotemporal transmission trends or evaluating the effectiveness of various NPIs. While our study investigated the role of initial outbreak locations, which is also important for understanding the transmission mechanism and beneficial for prevention and control.

(2) Most infectious-disease transmission models have been built at large spatial scales such as countries, provinces, states, and cities. While our study delved into a fine-grained intra-urban scale and integrated the real human mobility data between intra-urban regions, which can depict the transmission process more precisely.

(3) We proposed three indexes (i.e., the number of infected cases, the number of affected regions, and the spatial diffusion range) to evaluate the spatiotemporal transmission risk of infectious diseases from multiple perspectives, which can also be used in future studies and practice.

(4) We systematically simulated, evaluated, compared, and analyzed the transmission risk caused by each of the intra-urban regions as initial outbreak locations, and revealed the influential factors behind high-risk locations, which is of great importance for understanding the transmission mechanism and guiding prevention and control.

The main limitation of this study may be that we took only one city as the study case. If data of more cities can be obtained in the future, we would like to make experiments for multiple cities and test if those conclusions are tenable for different cities.

## Acknowledgement

This research is supported by the National Natural Science Foundation of China (No. 41901391, 41771441), Natural Science Foundation of Guangdong Province (No.2021A1515011191), and Major Science and Technology Projects of Xinjiang Uygur Autonomous Region (No. 2020A03004-4).